%% file: main.tex
\documentclass[
	reprint,
	amsmath,amssymb,
	prb,
	aps,
	superscriptaddress
]{revtex4-2}

\usepackage{natbib}
\usepackage{graphicx}
\usepackage{dcolumn}
\usepackage{bm}
\usepackage{physics}
\usepackage{amsmath}
\usepackage{nicefrac}
\usepackage{xcolor}
\usepackage{cancel}
\usepackage{microtype}
\usepackage{yquant}
\usepackage{quantikz}
\usepackage{array}

\newcommand\thiccboi[1]{\noalign{\hrule height #1}}

\begin{document}

\usetikzlibrary{shapes.misc}
\tikzset{cross/.style={cross out, draw=black, minimum size=2*(#1-\pgflinewidth), inner sep=0pt, outer sep=0pt},
cross/.default={2.5pt}}	

\title{Error-divisible two-qubit gates}

\author{David Rodr\'iguez P\'erez}
\email{drodriguezperez@mines.edu}
\author{Paul Varosy}
\affiliation{Department of Physics, Colorado School of Mines, Golden, CO 80401, USA}

\author{Ziqian Li}
\author{Tanay Roy}
\affiliation{James Franck Institute, University of Chicago, Chicago, IL 60637, USA}

\author{Eliot Kapit}
\affiliation{Department of Physics, Colorado School of Mines, Golden, CO 80401, USA}

\author{David Schuster}
\affiliation{James Franck Institute, University of Chicago, Chicago, IL 60637, USA}

\graphicspath{{figures/}}

\begin{abstract}	
	We introduce a simple, widely applicable formalism for designing ``error-divisible" two qubit gates: a quantum gate set where fractional rotations have proportionally reduced error compared to the full entangling gate. In current noisy intermediate-scale quantum (NISQ) algorithms, performance is largely constrained by error proliferation at high circuit depths, of which two-qubit gate error is generally the dominant contribution. Further, in many hardware implementations, arbitrary two qubit rotations must be composed from multiple two-qubit stock gates, further increasing error. This work introduces a set of criteria, and example waveforms and protocols to satisfy them, using superconducting qubits with tunable couplers for constructing continuous gate sets with significantly reduced error for small-angle rotations. If implemented at scale, NISQ algorithm performance would be significantly improved by our error-divisible gate protocols. 
\end{abstract}

\maketitle

\input{sections/introduction}
\input{sections/principles}
\input{sections/hardware}
\input{sections/vqe}
\input{sections/conclusions}
\input{sections/acknowledgements}

\bibliography{edg.bib}

\onecolumngrid
\appendix
\input{appendix/hamiltonianDerivation}
\input{appendix/data}

\input{appendix/circuits}

\end{document}

%% file: sections/introduction.tex
\section{\label{sec:1}Introduction}

Many advances in quantum hardware have focused on working towards fault-tolerant, universal quantum computation \cite{fowler_surface_2012,terhal_quantum_2015,barends_superconducting_2014}. However, the challenge of engineering a fully, error-correcting logical qubit is a formidable one, and advances in state coherence and gate errors need significant improvement before quantum computation reaches fault-tolerance. Nevertheless, current qubit implementations have reached coherence and control levels such that they can perform NISQ algorithms \cite{preskill_quantum_2018}, and may even be able to run tasks which surpass classical computers \cite{arute_quantum_2019}. NISQ variational algorithms, such as VQE and QAOA \cite{farhi_quantum_2014,harrigan_quantum_2021,peruzzo_variational_2014,mcclean_theory_2016}, have proven particularly useful in approximating the ground state energy of difficult Hamiltonians. However, the performance of these algorithms depends on the circuit depth that can be achieved, which in turn depends on qubit coherence and gate error.

The inaccuracy of modern quantum algorithm implementations is dominated by two-qubit gate errors, as single-qubit gate errors are typically an order of magnitude smaller \cite{majer_coupling_2007,yan_tunable_2018,xu_high-fidelity_2020,zhao_switchable_2020,collodo_implementation_2020,rasmussen_simple_2020}. This is compounded by the traditional approach of implementing arbitrary two-qubit gates using a stock gate set, in which the large numbers of two-qubit rotations that may be required by a variational algorithm could be decomposed into at most three CZ gates and additional single-qubit gates \cite{khaneja_cartan_2001}. While this gate decomposition is crucial for digital error correcting codes \cite{fowler_surface_2012}, it proliferates gate error for NISQ algorithms needing a more diverse two-qubit gate set. To address this, research in implementing continuous sets of gates that perform arbitrary two-qubit rotations \cite{foxen_demonstrating_2020,abrams_implementation_2020} has shown up to a 30\% reduction in gate depth \cite{peterson_fixed-depth_2020}.

This work presents error-divisible gates, which realizes smaller angle, two-qubit rotations that are prevalent in variational algorithms. The realization of smaller-angle rotations is done so at a corresponding smaller gate time---e.g. for full $\theta$ rotations at gate time $t_q$, $\theta/2$ rotations are done at $t_g/2$ time. This can provide an opportunity to execute even deeper circuits given the further reduction in energy loss by virtue of having shorter gates.

\subsection{\label{sec:1-errorDivisibility}Error divisibility}

\begin{figure}
	\includegraphics[width = 0.49\textwidth]{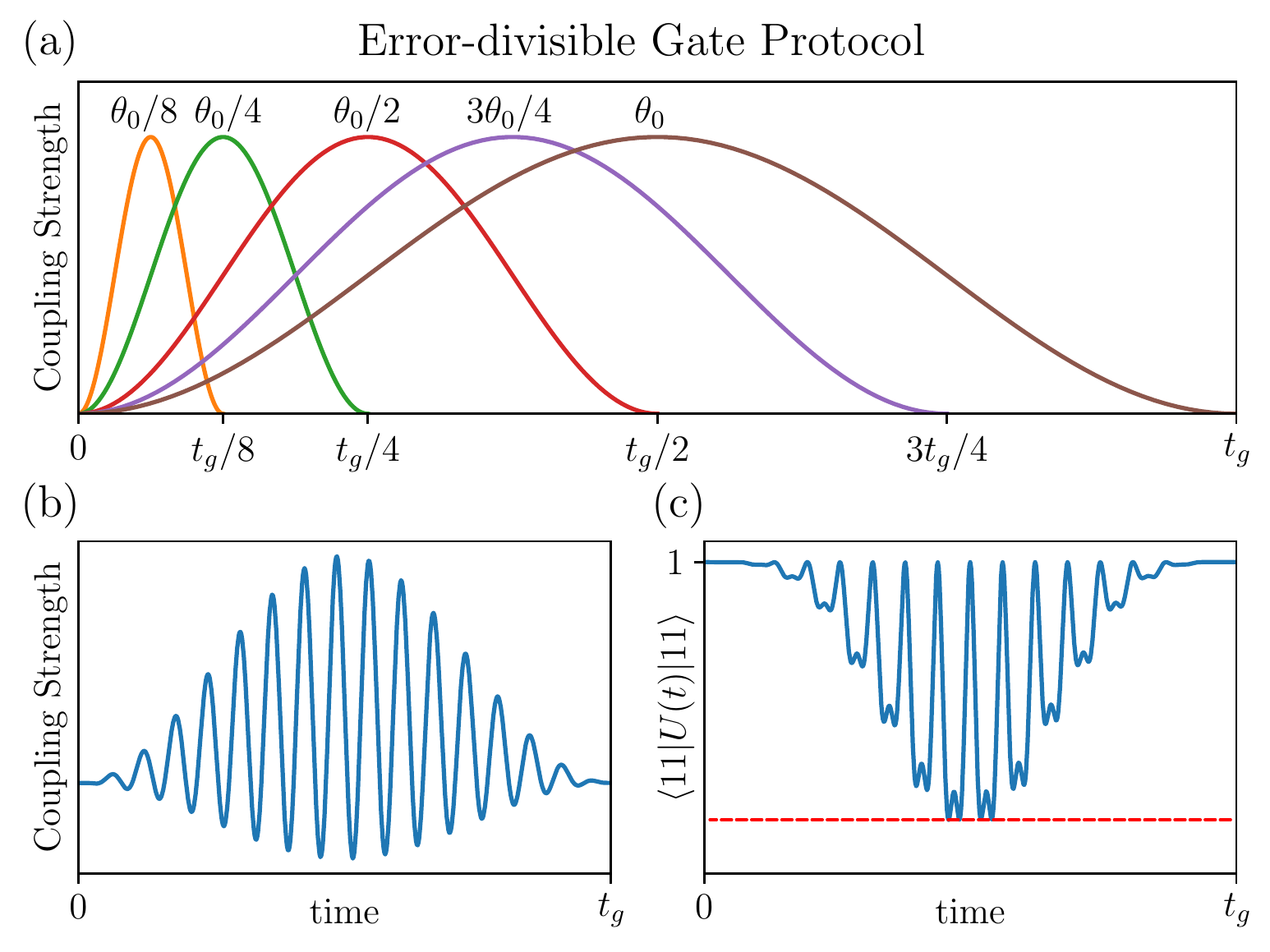}
	\caption{\label{fig:edg_demo}Illustration of our proposed error divisibility protocol. (a) Two-qubit rotation $\theta_0$ at gate time $t_g$, along with fractional rotations $\theta_0/n$ at corresponding gate times $t_g/n$. Note that these waveforms are only meant to demonstrate the idea of error-divisible fractional rotations at fractional gate times, and do not represent actual waveforms (amplitudes and frequencies---shown in (b)---can be different for each partial gate). (b) Example gate strength waveform for achieving error divisibility using a Gaussian gate envelope superimposed with a fast-oscillating counterterm. (c) Corresponding probability of gate $U(t)$ inducing transitions outside of $\ket{11}$ throughout the gate evolution in (b), where $U(t) = T\exp\left(-2\pi i\int_0^{t_g}\Omega(t)\left(\ketbra{01}{10} + \ketbra{10}{01} \right)dt\right)$ with coupling strength $\Omega(t)$ like in (b). An ideal exchange operation would preserve unity throughout the gate, but accounting for nonlinear leakage states, we see off-resonant transitions outside of $\ket{11}$, creating these dips in (c), the maximum indicated by the red-dashed. In typical operating regimes, the coupling strength in (b) is on the order of tens of MHz, $t_g \approx 30$ ns, and the maximum dip from unity---which varies by the target rotation---can be $<10^{-2}$ for appropriate parameter choices.}
\end{figure}

The basic idea for error divisibility is to implement a small two-qubit gate rotation without using multiple larger-rotation gates by instead proposing waveforms that are pulse-shaped to perform fractional rotations at a corresponding fractional gate time. This is illustrated in Fig.~\ref{fig:edg_demo}(a), where a full two-qubit rotation $\theta_0$ is achieved a gate time $t_g$, and fractional rotations $\theta_0/n$ are run with corresponding gate times $t_g/n$. The full gate time $t_g$ is chosen as small as possible so that the rotation $\theta_0$ is executed at a chosen, acceptable intrinsic error rate (not considering qubit decoherence). The fractional two-qubit gates implemented with the proposed protocol do not increase intrinsic qubit error significantly. 

To address the difficulty in pulse-shaping gates of variable length for different target rotations, this work borrows ideas from \cite{xiong_arbitrary_2021} and \cite{perez_improved_2020}, in which off-resonant energetics are used to suppress dispersive shift effects. We propose simple waveforms that are superimposed Gaussian-like wave envelopes with fast-oscillating counterterms, as shown in Fig.~\ref{fig:edg_demo}(b). These waveforms do not require a high level of fine-tuning, and significant errors are not introduced from small variations in the final waveform. The fast-oscillating counterterm serves to suppress intrinsic gate error throughout the duration of the gate. Namely, the ones this work focuses on are leakage into higher-energy excited states, and dynamic stray $ZZ$ interactions that can create an effective partial CPHASE. These processes are the result of off-resonant mixing throughout the gate, which can be mitigated by additional off-resonant processes---realized as the fast-oscillating counterterms. This is illustrated in Fig.~\ref{fig:edg_demo}(c), where suppression of the gate error through the duration of the gate can be seen as a result of the oscillating waveform. The dashed red line indicates the maximum gate error, which can be reduced through minimal fine tuning of the waveform's small parameter space.

While the work presented here completely ignores random qubit error like amplitude dampening and dephasing, these are automatically addressed by the fact that this protocol shortens the gate time for each smaller rotation angle, allowing less time for decoherence. This provides an even bigger reduction in the error per gate than other continuous gate implementations \cite{foxen_demonstrating_2020,abrams_implementation_2020}, allowing for a greater circuit depth for NISQ algorithms without a significant increase in gate complexity.

%% file: sections/principles.tex
\section{\label{sec:2}Principles of implementation}

\begin{figure*}
	\includegraphics[width = \textwidth]{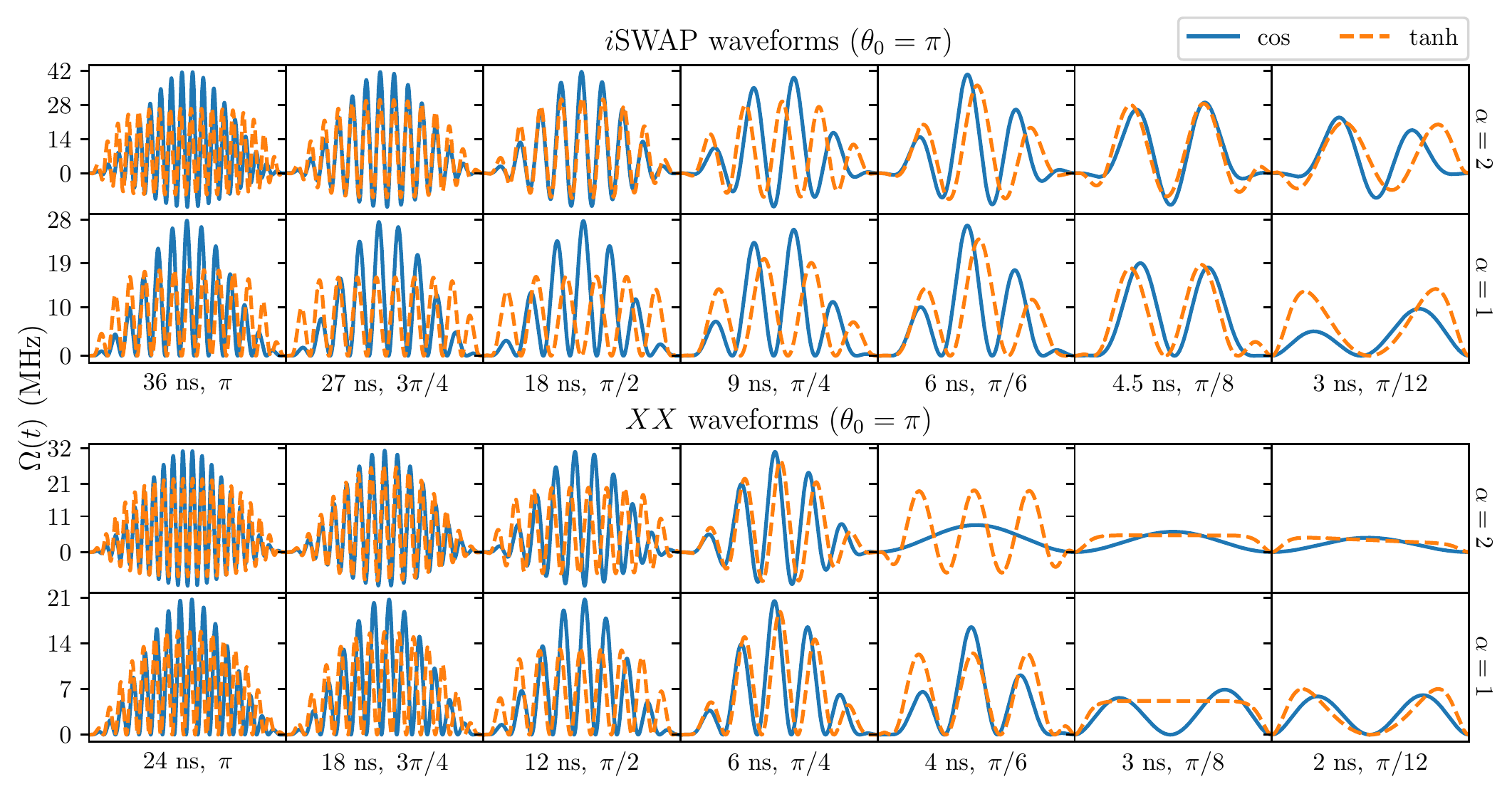}
	\caption{\label{fig:edg_pulses}Waveforms given by Eq.~\ref{eq:envelopes} for different rotations of $i$SWAP (top) and $XX$ (bottom) gates. Waveforms for Eq.~\ref{eq:envelopes}a and Eq.~\ref{eq:envelopes}b are labelled simply as ``cos'' and ``tanh'', respectively, in the legend. The first row of both $i$SWAP and $XX$ waveforms use $\alpha = 2$ in Eq.~\ref{eq:couplingStrength}, while the bottom rows use $\alpha = 1$ to restrict the waveforms' positivity. The full rotation waveforms are on the far left, subsequent plots reduce the target rotation and corresponding gate time by the fractional series $\{3/4,\ 1/2,\ 1/4,\ 1/6,\ 1/8,\ 1/12\}$.}
\end{figure*}

Consider two transmons with energies $\omega_1$ and $\omega_2$ with nonlinearity $-\delta$, coupled by the term $g(t)\left(a_1 + a_1^{\dagger}\right)\left(a_2 + a_2^{\dagger}\right)$. With appropriate choices of $g(t)$, we can resonantly drive a photon exchange between the two qubits. Choosing $g(t) = g_0\Omega(t)\cos[2\pi(\omega_1 - \omega_2)]$ gives us the rotating frame Hamiltonian (Appendix~\ref{app:Hamiltonian})

\begin{equation}\label{eq:RFHam}
	\begin{split}
		H = &\frac{\delta}{2}\left( a_1^{\dagger}a_1^{\dagger}a_1a_1 + a_2^{\dagger}a_2^{\dagger}a_2a_2 \right)\\
		&- \Omega(t)g_0\left(a_1^{\dagger}a_2 + a_1a_2^{\dagger}\right).
	\end{split}
\end{equation}
By going to second order in perturbation theory and eliminating the second excited energy state of the transmons, the qubit Hamiltonian becomes
\begin{equation}\label{eq:perturbationHamiltonian}
	\begin{split}
		H \simeq -&\Omega(t)g_0\left( \sigma_1^+\sigma_2^- + \sigma_1^-\sigma_2^+ \right)\\
		&+ \frac{\Omega(t)^2g_0^2}{\delta}(1+\sigma_1^z)(1+\sigma_2^z),		
	\end{split}
\end{equation}
where we can see an effective dispersive shift term that applies a $\sigma_1^z\sigma_2^z$ term concurrently with the exchange Hamiltonian, $H = H_{\textrm{ex}} + H_{ZZ}$. To eliminate this dispersive shift, we let $\Omega(t)$ take on a shape like that of Fig.~\ref{fig:edg_demo}(b),
\begin{equation}\label{eq:couplingStrength}
	\Omega(t) = \Omega_0(t)\left[1 + \alpha\sin(2\pi ft)\right],
\end{equation}
where $\Omega_0(t)$ is a slow-evolving gate envelope, and the fast-oscillating counterterm is determined by a frequency $f$, and $\alpha$ determines whether the waveform is positive valued---a condition required by some hardware implementations \cite{chen_qubit_2014}. Supposing that the slow-evolving envelope in Eq.~\ref{eq:couplingStrength} is red-detuned from the off-resonant transition causing phase accumulation, the fast-oscillating term adds a corresponding dispersive shift blue-detuned term with opposite sign, and can cancel it out. Examples of engineering these dispersive shifts from off-resonant processes is given in \cite{kapit_error-transparent_2018}. This is a similar approach to what was discovered in \cite{perez_improved_2020}, where an oscillatory offset proportional to the nonlinearity was found to suppress population of the leakage states.

The examples presented here focus on exploring different wave envelopes $\Omega_0(t)$ for two different, candidate gate systems---$i$SWAP$(\theta)=\exp[i\theta/2\left(\sigma^+_1\sigma^-_2 + \sigma^-_1\sigma^+_2\right)]$ and $X$C$X$$(\theta) = \exp[i\theta/4\left(1 + \sigma_1^x\right)\left(1 + \sigma_2^x\right)]$, where a full gate rotation $\theta_0$ (like the illustration from Fig.~\ref{fig:edg_demo}) would be $\theta_0 = \pi$. The $X$C$X$$(\theta)$ gate set---the $x$-basis version of CPHASE$(\theta)$---is chosen as it is easier to engineer than CPHASE. Also note that $X$C$X$ reduces to single-qubit operations and one two-qubit $XX$ operation, which along with $i$SWAP, becomes the focus of the following example waveforms.

\subsection{\label{sec:2-results}Example Waveforms}

\begin{figure}
	\includegraphics[width = 0.485\textwidth]{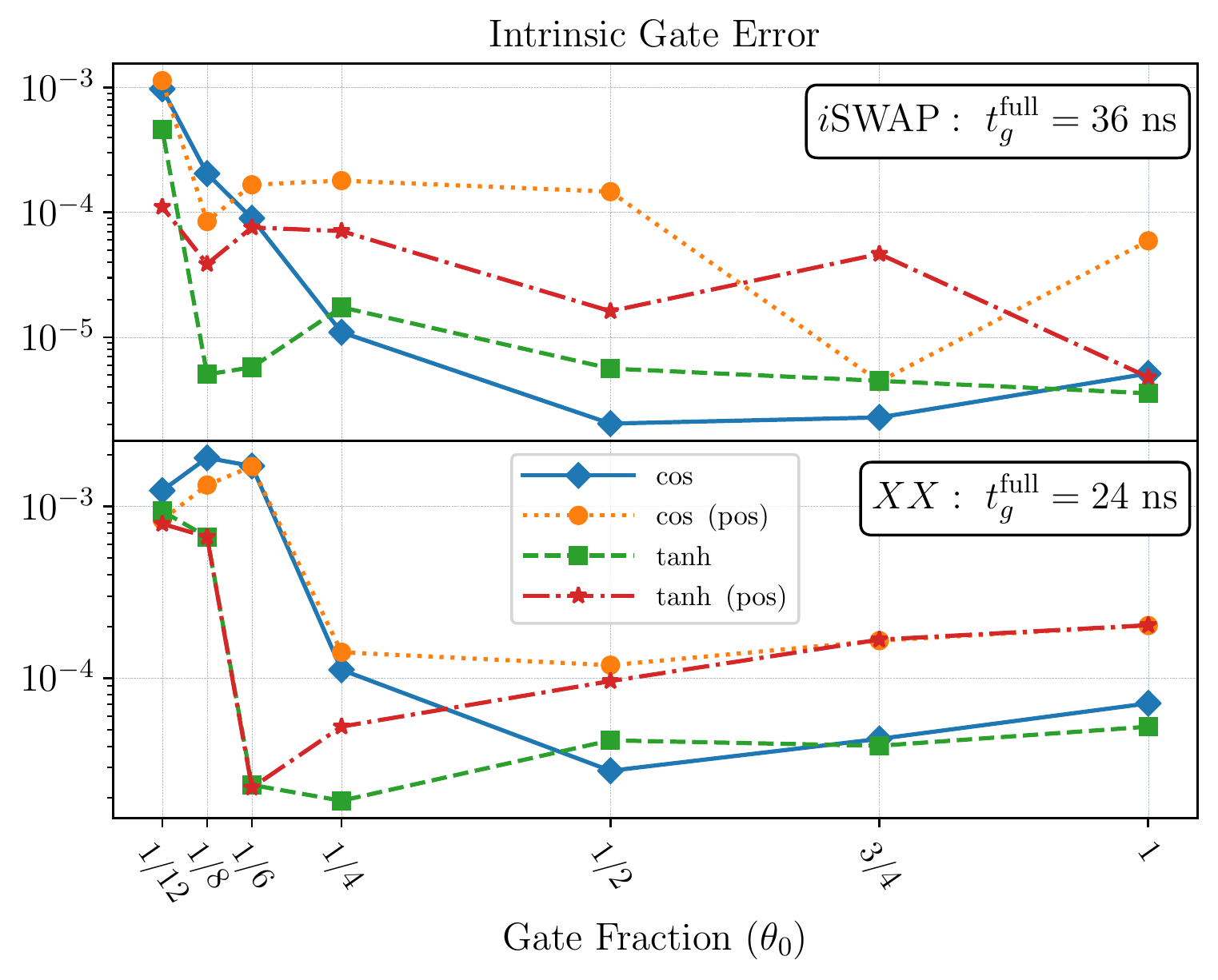}
	\caption{\label{fig:edg_results}Intrinsic gate error results for fractional gate rotations for iSWAP (top, 36 ns gate for full $\pi$ rotation) and $XX$ (bottom, 24 ns for full $\pi$ rotation) gates using both positive-definite (pos) and freely oscillating gate envelopes Eq.~\ref{eq:envelopes}a (cos) and Eq.~\ref{eq:envelopes}b (tanh). For fractional iSWAP gates, we achieve a gate error $< 10^{-4}$ down to $1/8^{\textrm{th}}$ of a gate using tanh envelopes and $1/6^{\textrm{th}}$ for $XX$ gates.}
\end{figure}

We propose some example waveforms with the characteristic fast-oscillating counterterm, as shown in Fig.~\ref{fig:edg_demo}(b). We focus on two different options for the slow-evolving envelope $\Omega_0(t)$ in Eq.~\ref{eq:couplingStrength}:
\begin{subequations}\label{eq:envelopes}
	\begin{align}
		&\Omega_0(t) = \frac{A}{2}\left[1 - \cos\left(2\pi\frac{t}{t_g}\right)\right],\ \textrm{and}\\
		\begin{split}
		\Omega_0(t) = &A\Bigg[\tanh\left(\gamma\frac{t}{t_g}\right)\\
		&- \tanh\left(\gamma\left[\frac{t}{t_g} - 1\right]\right) - \tanh\gamma\Bigg]^2.
		\end{split}
	\end{align}
\end{subequations}
For both Eq.~\ref{eq:envelopes}a and Eq.~\ref{eq:envelopes}b, the fast-oscillating part takes the form from Eq.~\ref{eq:couplingStrength}, where we numerically tune the frequency $f$ for different target rotations, and choose $\alpha = 1$ for a positive-definite waveform, and $\alpha = 2$ for waveforms without such restrictions. $A$ and $\gamma$ are also numerically tuned for each target rotation, leaving only 2 parameters to tune for Eq.~\ref{eq:envelopes}a waveforms, and three for Eq.~\ref{eq:envelopes}b. These are shown in Fig.~\ref{fig:edg_pulses} for $i$SWAP and $XX$ gates, using nonlinearity $\delta = 300$~MHz. Per the protocol laid out in Fig.~\ref{fig:edg_demo}, we first find a waveform for a full two-qubit rotation with a minimum gate time $t_g$ while maintaining an acceptably low intrinsic error rate. We can then find waveforms for fractional rotations with corresponding fractional gate times, tuning the parameters $A,\ \gamma$, and $f$ to maintain a low gate error. The results presented here do not consider decoherence and only account for leakage. 

Error rate results for these fractional gates are shown in Fig.~\ref{fig:edg_results}. We can see that error-divisibility is achieved for both $i$SWAP and $XX$ gates down to 1/8th and 1/6th, respectively, using the profile Eq.~\ref{eq:envelopes}b---this profile gives the best results. Considering that the target angle for a full $X$C$X$ operation is half on an $i$SWAP, we expect this error-divisible scheme to break down faster for fractional $XX$ operations, where small $t_g$ starts to approach $1/\delta \approx 3\ \text{ns}$. The limits of this scheme are evidenced by the noticeable increases in gate error down to $1/12$ of a rotation, shown in Fig.~\ref{fig:edg_results}, where we also point out a general higher gate error using positive-definite waveforms. We also note that intrinsic gate error can be further reduced using higher resolution in pulse-shaping with a fourier series waveform to arbitrary precision. But this would require an arbitrary number of parameters to calibrate, making this option less-desirable. The waveforms in Eq.~\ref{eq:envelopes} provide a simple parameter space to calibrate that matches experimental hardware. The complete set of criteria we have proposed pave a way towards greater depth circuits. Results and waveforms shown in Fig.~\ref{fig:edg_pulses} and Fig.~\ref{fig:edg_results} were obtained using unitary dynamics (ignoring energy loss and dephasing), and parameters were found using differential evolution methods \cite{feldt_robertfeldtblackboxoptimjl_2021}.


%% file: sections/hardware.tex
\section{\label{sec:3}Hardware implementations}

In the current state of superconducting hardware, advances in optimal control methods have mostly eliminated single-qubit gate errors \cite{motzoi_simple_2009,mottonen_high-fidelity_2006,safaei_optimized_2009,steffen_accurate_2003}. However, multiple-qubit gate errors are still an order of magnitude greater than single-qubit gates \cite{majer_coupling_2007,yan_tunable_2018,xu_high-fidelity_2020,zhao_switchable_2020,collodo_implementation_2020,rasmussen_simple_2020}, suffering from a combination of random qubit error during the gate, crosstalk \cite{mundada_suppression_2019,sheldon_procedure_2016,winick_simulating_2020}, and calibration drift over time in systems using tunable architectures \cite{machnes_tunable_2018,stehlik_tunable_2021,dai_calibration_2021}. This error-divisible scheme intrinsically addresses random qubit error by reducing the duration of the gates in NISQ algorithms requiring smaller two-qubit rotations. This can be reliably done with our proposed scheme provided there are not error-limiting steps. This means gate protocols using tunable qubit energies \cite{dicarlo_demonstration_2009,barends_coherent_2013}, where coupling is done by bringing them into resonance with each other, cannot be considered for this protocol due to the fixed amount of time necessary for tuning the qubit energies. The presence of fixed amounts of time also eliminates cross-resonance gates \cite{paraoanu_microwave-induced_2006,malekakhlagh_first-principles_2020,rigetti_fully_2010,magesan_effective_2020}, where there is a minimum duration needed to suppress leakage in the process of driving qubits into higher excited energy states.

This leaves us with qubit architectures with fixed qubit energies \cite{mckay_universal_2016,caldwell_parametrically_2018}, in which coupling is achieved by driving a coupler at the appropriate frequencies. These allow the implementation of the waveforms presented in Fig.~\ref{fig:edg_pulses}, creating a path towards error-divisibility. The ability to realize this protocol with any system using a tunable coupler circuit allows for an easy generalization to qubits with large and small nonlinearities. Thus, while the analysis and simulations done in this work assumes a transmon architecture, it is easily realizable for flux qubits \cite{orlando_superconducting_1999,mooij_josephson_1999,nguyen_high-coherence_2019} and fluxoniums \cite{pop_coherent_2014,manucharyan_fluxonium_2009}. 

%% file: sections/vqe.tex
\section{\label{sec:4}VQE example}

\begin{figure}
	\vspace{0.8cm}
	\begin{center}
	\begin{tikzpicture}
		\begin{yquant}[register/minimum height=5.35mm]
			qubit {$\ket{q_{\idx}}$} q[6];
			swap (q[0], q[1]);
			swap (q[2], q[3]);
			swap (q[4], q[5]);

			zz (q[0], q[1]);
			zz (q[2], q[3]);
			zz (q[4], q[5]);

			swap (q[1], q[2]);
			swap (q[3], q[4]);

			zz (q[1], q[2]);
			zz (q[3], q[4]);

			box {$R_Z(\phi)$} q[0, 2, 4];
		\end{yquant}
		\node at (1.1,-0.57) {$\theta$};
		\node at (1.1,-1.84) {$\theta$};
		\node at (1.1,-3.11) {$\theta$};
		\node at (2.3,-1.2) {$\theta$};
		\node at (2.3,-2.48) {$\theta$};

		\node at (0.5,-0.57) {$\theta$};
		\node at (0.5,-1.84) {$\theta$};
		\node at (0.5,-3.11) {$\theta$};
		\node at (1.7,-1.2) {$\theta$};
		\node at (1.7,-2.48) {$\theta$};

		\draw[fill=white] (0.35 ,-0.26) circle (3pt);
		\draw (0.35 ,-0.26) node[cross] {};
		\draw[fill=white] (0.35 ,-0.89) circle (3pt);
		\draw (0.35 ,-0.89) node[cross] {};
		\draw[fill=white] (0.35 ,-1.53) circle (3pt);
		\draw (0.35 ,-1.53) node[cross] {};
		\draw[fill=white] (0.35 ,-2.16) circle (3pt);
		\draw (0.35 ,-2.16) node[cross] {};
		\draw[fill=white] (0.35 ,-2.8) circle (3pt);
		\draw (0.35 ,-2.8) node[cross] {};
		\draw[fill=white] (0.35 ,-3.44) circle (3pt);
		\draw (0.35 ,-3.44) node[cross] {};
		\draw[fill=white] (1.55 ,-0.89) circle (3pt);
		\draw (1.55 ,-0.89) node[cross] {};
		\draw[fill=white] (1.55 ,-1.53) circle (3pt);
		\draw (1.55 ,-1.53) node[cross] {};
		\draw[fill=white] (1.55 ,-2.16) circle (3pt);
		\draw (1.55 ,-2.16) node[cross] {};
		\draw[fill=white] (1.55 ,-2.8) circle (3pt);
		\draw (1.55 ,-2.8) node[cross] {};
	\end{tikzpicture}
	\end{center}
	\caption{\label{fig:vqe_cirq}Circuit diagram for a single layer of VQE implementing the Hamiltonian Eq.~\ref{eq:AFM} using six qubits. The ``$\theta$'' labels next to the $i\text{SWAP}$ and CPHASE gates denote their partial rotations by angle $\theta$.} 
\end{figure}
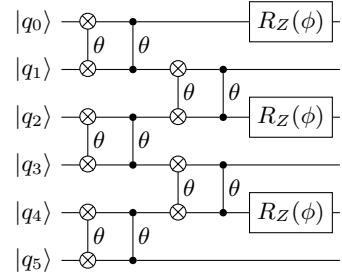

\begin{figure}
	\includegraphics[width = 0.485\textwidth]{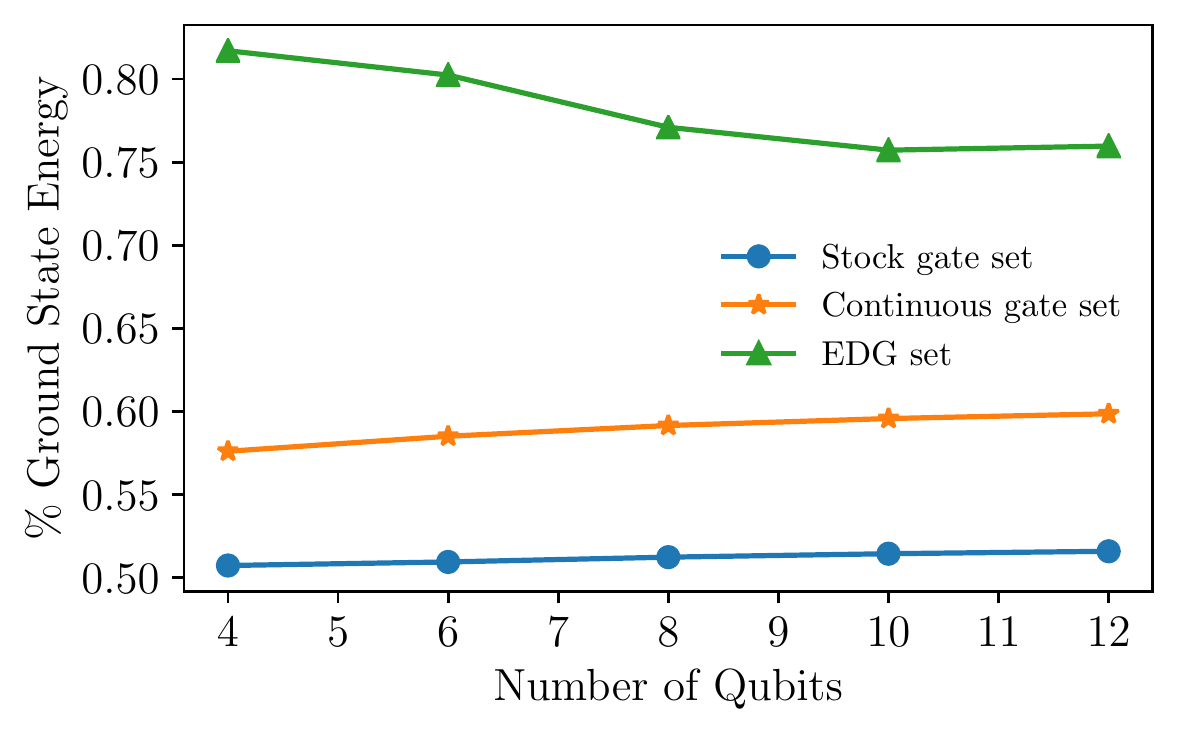}
	\caption{\label{fig:vqe_results}Comparisons for VQE fraction of ground state results using a stock gate set (blue square), continuous gates (orange star), and the proposed error-divisible gate scheme (green triangle).}
\end{figure}

To further motivate the use of the error-divisible protocol proposed in this work, we demonstrate the utility of running a variational algorithm with access to error-divisible gates. The example problem we consider is an adiabatic evolution
\begin{equation}\label{eq:adiabatic}
	H(t) = \left(1- \frac{t}{T}\right)H_{\text{pin}} + \frac{t}{T}H_{\text{prob}},
\end{equation}
where $T$ is the total runtime, $H_{\text{pin}}$ is a pinning Hamiltonian with an easily solvable solution, and $H_{\text{prob}}$ is a harder problem Hamiltonian whose ground state we are trying to determine by starting in the ground state of $H_{\text{pin}}$ and adiabatically evolving $H(t)$. Approximating with a Trotter decomposition \cite{beig_finding_2005}
\begin{equation}\label{eq:trotter}
	e^{-i\,dt(A + B)} = e^{-i\,dtA}e^{-i\,dtB} + O(dt^2),
\end{equation}
we are able to implement the Hamiltonian as a collection of one- and two-qubit gates on hardware. Letting $N$ be the number of Trotter layers, we set $dt = T/N$.

We choose to find the ground state of an Antiferromagnetic Heisenberg Model (AFM), letting the pinning and problem Hamiltonians in Eq.~\ref{eq:adiabatic} be
\begin{subequations}\label{eq:AFM}
	\begin{align}
		&H_{\textrm{pin}} = V\sum_{j\ \text{even}} \sigma_j^z\\
		H_{\text{prob}} = J\sum_{j = 0}^{n_q-2}&\left(\sigma_j^x\sigma_{j+1}^x + \sigma_j^y\sigma_{j+1}^y + \sigma_j^z\sigma_{j+1}^z\right),
	\end{align}
\end{subequations}
where $n_q$ is the number of qubits, and $V$ and $J$ are energy scales, which we set $V=J=1$ here. The circuit for Eq.~\ref{eq:AFM} is shown in Fig.~\ref{fig:vqe_cirq}, where the $Z$ rotations correspond to the pinning Hamiltonian Eq.~\ref{eq:AFM}a, and the $i$SWAP$(\theta)$ CPHASE$(\theta)$ sequence performs the problem Hamiltonian Eq.~\ref{eq:AFM}b by the approximation Eq.~\ref{eq:trotter}. $T$ and $N$ are optimized after every layer in Fig.~\ref{fig:vqe_cirq} using a nonlinear optimization routine \cite{kaelo_variants_2006,johnson_stevengjnlopt_2021}.

The primary error model considered here is a 1\% depolarizing noise error rate for two-qubit gates and 0.1\% for single-qubit gates. We compare the performance of this simulated variational algorithm using error-divisible gates, continuous gates \cite{abrams_implementation_2020,foxen_demonstrating_2020}, and a stock gate set. Results are shown in Fig.~\ref{fig:vqe_results} with, $\theta = 2J(t/N)\,dt$ and $\phi = 2V(1-t/T)\,dt$. An implementation of Fig.~\ref{fig:vqe_cirq} using a stock gate set would require two two-qubit gates and at least 4 single-qubit gates (Appendix~\ref{app:circuits}), giving an error rate of at least 2.4\% per two-qubit operation in the VQE layer. With access to continuous gates, there would be no need for such decompositions, and thus the two-qubit gate error rate is the 1\% we define. Using error-divisible gates, implementing fractional gates with fractional gate times, results in a corresponding fractional error rate, and is thus determined by $N$. These clear advantages are demonstrated in simulation as seen in Fig.~\ref{fig:vqe_results}, providing a clear example of the potential benefit in implementing error-divisible gates for near-term quantum computers.

%% file: sections/conclusions.tex
\section{Conclusions}

We have provided a set of criteria for implementing error-divisible two-qubit gates using currently available technology. We introduced the notion of fractional gates for which error-divisibility provides a huge advantage for NISQ algorithms. At the heart of this criteria is the ability to generate a fast-oscillating waveform that can cancel dispersive shift effects. To that end, we explored two sets of wave envelopes superimposed with fast-oscillating terms on two family of gates---$i$SWAP$(\theta)$ and $XX(\theta)$, and found error-divisibility down to 1/6th and 1/8th of a full rotation, respectively. We then further motivated error-divisible gates by showcasing their potential advantage by simulating a variational algorithm for the adiabatic evolution of an antiferromagnetic Heisenberg model problem Hamiltonian.

%% file: sections/acknowledgements.tex
\section{Acknowledgements}

We would like to thank Eric Jones and Zhijie Tang for useful discussions. This work was supported by the NSF grant (PHY-1653820) and ARO grant No. W911NF-18-1-0125. Simulations for results in section~\ref{sec:4} were obtained using the QuEST quantum simulator \cite{jones_quest_2019}. Both sets of results in sections~\ref{sec:2-results} and \ref{sec:4} were obtained using the Wendian high performance computer at Colorado School of Mines.

%% file: appendix/hamiltonianDerivation.tex
\section{Hamiltonian Derivation}\label{app:Hamiltonian}

We explicitly derive the rotating frame Hamiltonian (Eq.~\ref{eq:RFHam}) for the two coupled qubits. Starting from the two coupled qubits with energies $\omega_1$, $\omega_2$, and nonlinearity $\delta$, the system is described by the Duffing oscillator Hamiltonian in the energy basis as well as a coupling Hamiltonian,

\begin{equation}\label{eq:duffingHamiltonian}
	H = \omega_1 \hat{n}_1 + \omega_2 \hat{n}_2 - \frac{\delta}{2}\left( a^{\dagger}_1a^{\dagger}_1 a_1a_1 + a^{\dagger}_2a^{\dagger}_2 a_2a_2 \right) + g(t)\left(a_1^{\dagger} + a_1 \right) \left(a^{\dagger}_2 + a_2 \right),
\end{equation}
with $\hat{n}_j = a_j^{\dagger}a_j$, $g(t) = g_0A(t)\cos(2\pi(\omega_1 - \omega_2)t)$, and we have let $\hbar=1$. Recognizing this Hamiltonian with a time-independent and time-dependent parts, $H(t) = H_0 + V(t)$, this obeys the Schr\"odinger equation $i\partial_t\ket{\psi(t)} = H(t)\ket{\psi(t)}$. We apply a unitary rotation transformation $U_R(t)$ such that a new state can be defined
\begin{equation}\label{eq:rotationState}
	\ket{\phi(t)} = U_R(t)\ket{\psi(t)}.
\end{equation}
Looking at the time evolution of $\ket{\phi(t)}$,
\begin{equation}
	\begin{split}
		i\partial_t\ket{\phi(t)} &= i\partial_t\left(U_R(t)\ket{\psi(t)}\right)\\
								 &= i\partial_tU_R(t)\ket{\psi(t)} + U_R(t)(i\partial_t\ket{\psi(t)})\\
								 &= i\dot{U}_R(t)\ket{\psi(t)} + U_R(t)H(t)\ket{\psi(t)}\\
								 &= i\dot{U}_R(t)U_R^{\dagger}(t)\ket{\phi(t)} + U_R(t)H(t)U_R^{\dagger}(t)\ket{\phi(t)},
	\end{split}
\end{equation}
where we have used \ref{eq:rotationState} to replace $\ket{\psi(t)} = U_R^{\dagger}(t)\ket{\phi(t)}$. From here, it's clear that the state $\ket{\phi(t)}$ also obeys the Schr\"odinger equation with a modified Hamiltonian
\begin{equation}
	i\partial_t\ket{\phi(t)} = \left( i\dot{U}_R(t)U_R^{\dagger}(t) + U_R(t)H(t)U_R^{\dagger}(t) \right)\ket{\phi(t)}.
\end{equation}
This is our rotating frame Hamiltonian $i\partial_t\ket{\phi(t)} = H_R(t)\ket{\phi(t)}$. Separating the time dependent part of the Hamiltonian \ref{eq:duffingHamiltonian}, we express $H_R(t)$ as
\begin{equation}
	H_R(t) = i\dot{U}_R(t)U_R^{\dagger}(t) + U_R(t)H_0U_R^{\dagger}(t) + U_R(t)V(t)U_R^{\dagger}(t),
\end{equation}
allowing us to separate this problem into solving three smaller parts. Letting $U_R(t) = e^{i(\omega_1\hat{n}_1 + \omega_2\hat{n}_2)t}$, we have
\begin{subequations}\label{eq:hamiltonianParts}
	\begin{align}
		i\dot{U}_R(t)U_R^{\dagger}(t) &= -(\omega_1\hat{n}_1 + \omega_2\hat{n}_2)\\
		U_R(t)H_0U_R^{\dagger}(t) &= \omega_1\hat{n}_1 + \omega_2\hat{n}_2 - \frac{\delta}{2}\left(a_1^{\dagger}a_1^{\dagger}a_1a_1 + a_2^{\dagger}a_2^{\dagger}a_2a_2\right)\\
		U_R(t)V(t)U_R^{\dagger}(t) &= -g(t)\left( a_1^{\dagger}e^{i\omega_1t} + a_1e^{-i\omega_1t} \right)\left( a_2^{\dagger}e^{i\omega_2t} + a_2e^{-i\omega_2t} \right)
	\end{align}
\end{subequations}
where we have used the relations $a^{\dagger}e^{-i\omega\hat{n}t} = e^{i\omega\hat{n}t}a^{\dagger}e^{i\omega t}$ and $ae^{-i\omega\hat{n}t} = e^{i\omega\hat{n}t}ae^{-i\omega t}$. Replacing $g(t)$ (ignoring $2\pi$ and using $\cos\left[(\omega_1 - \omega_2)t\right]$ for the moment) and using Euler's formula, \ref{eq:hamiltonianParts}c becomes

\begin{equation}\label{eq:RWA}
	\begin{split}
		U_R(t)V(t)U_R^{\dagger}(t) = -\frac{g_0}{2}A(t)\Big( &a_1^{\dagger}a_2^{\dagger}e^{2i\omega_1t} + a_1^{\dagger}a_2e^{2i(\omega_1 - \omega_2)t} + a_1a_2^{\dagger}\cancelto{\scriptstyle 1}{e^0} + a_1a_2e^{-i2\omega_2t}\\
		&+ a_1^{\dagger}a_2^{\dagger}e^{2i\omega_1t} + a_1^{\dagger}a_2\cancelto{\scriptstyle 1}{e^0} +a_1a_2^{\dagger}e^{-2i(\omega_1 - \omega_2)t} + a_1a_2e^{-2i\omega_1t} \Big).
	\end{split}
\end{equation}
Making the rotating wave approximation, we can toss all fast-oscillating terms and keep slow-oscillating and stationary ones. Note that if $\omega_1$ and $\omega_2$ are close or far apart only changes the final result by a factor of 2 under the approximation, since the terms oscillating at $(\omega_1 - \omega_2)$ are the same as the stationary ones. Putting together \ref{eq:hamiltonianParts} and \ref{eq:RWA}, our approximated, rotating frame Hamiltonian is
\begin{equation}\label{eq:finalRotatingFrameHam}
	H_{\textrm{RF}} = -\frac{\delta}{2}\left( a_1^{\dagger}a_1^{\dagger}a_1a_1 + a_2^{\dagger}a_2^{\dagger}a_2a_2 \right) - A(t)g_0\left(a_1^{\dagger}a_2 + a_1a_2^{\dagger}\right).
\end{equation}
Letting $H_{RF} = H_0 + H_1$, with $H_0 = \frac{\delta}{2}\left(a_1^{\dagger}a_1^{\dagger}a_1a_1 + a_2^{\dagger}a_2^{\dagger}a_2a_2\right)$ and $H_1 = - A(t)g_0\left(a_1^{\dagger}a_2 + a_1a_2^{\dagger}\right)$, we perturbatively find the effective Hamiltonian, treating $A(t)$ as fixed for the moment. Ignoring states of higher energy than $\ket{2}$, the eigensystem for $H_0$ is given by
\begin{equation}\label{eq:H0eigensystem}
	0\{\ket{0_10_2}, \ket{0_11_2}, \ket{1_10_2}, \ket{1_11_2}\},  -\delta\{\ket{0_12_2}, \ket{2_10_2}, \ket{1_12_2}, \ket{2_11_2}\}, -2\delta\ket{2_12_2}.
\end{equation}
Since the action of $H_1$ on any $\ket{2}$ state results in a $\ket{3}$ state, we ignore any corrections on those and focus on perturbative corrections for the $\ket{0}, \ket{1}$ subspace (0 eigenenergy states in \ref{eq:H0eigensystem}). Because these are degenerate, we diagonalize the perturbing Hamiltonian $H_1$ in this unperturbed eigenstate $\ket{n^0}$ subspace. Building out the $\bra{n_i^0}H_1\ket{n_j^0}$ matrix, we get
\begin{equation}
	\begin{pmatrix}
		0 & 0 & 0 & 0\\
		0 & 0 & -A(t)g_0 & 0\\
		0 & -A(t)g_0 & 0 & 0\\
		0 & 0 & 0 & 0
	\end{pmatrix},
\end{equation}
giving us the first-order corrections to the degenerate states
\begin{equation}\label{eq:firstCorrection}
	0(\ket{0_10_2}, \ket{1_11_2}), \pm\frac{A(t)g_0}{\sqrt{2}}(\ket{1_10_2} \mp \ket{0_11_2}).
\end{equation}
The only state in this subspace with a non-zero, second order correction is $\ket{1_11_2}$, yielding $\ket{0_12_2}$ and $\ket{2_10_2}$ upon the action of $H_1$, giving us
\begin{equation}\label{eq:secondCorrection}
	\frac{4A(t)^2g_0^2}{\delta}\ket{1_11_2}.
\end{equation}
Eliminating second-excited energy states now $\left(a^{\dagger}\rightarrow\sigma^+,\ a\rightarrow\sigma^-\right)$, and up to second order in perturbation theory---combining \ref{eq:firstCorrection} and \ref{eq:secondCorrection}, we get the effective Hamiltonian
\begin{equation}\label{eq:effectiveH}
	H \simeq -A(t)g_0\left( \sigma_1^+\sigma_2^- + \sigma_1^-\sigma_2^+ \right) + \frac{A(t)^2g_0^2}{\delta}(1+\sigma_1^z)(1+\sigma_2^z).
\end{equation}

This dispersive shift gives an effective C$Z$ and partial $i$SWAP. To counter this, we can choose $A(t) = A_0(t)[1 + B\sin(2\pi ft)]$, where $A_0(t)$ is a slow-evolving wave envelope. Given $f \gg g_0$, the leading exchange term in \ref{eq:effectiveH} remains unaffected, while the oscillatory effects create off-resonant dispersive shifts that counter the effects of the $\sigma^z_1\sigma^z_2$ term.

%% file: appendix/data.tex
\section{Simulation Data}\label{app:data}

Simulated data for waveforms and results in Fig.~\ref{fig:edg_pulses} and Fig.~\ref{fig:edg_results}. This set of data was obtained by numerically tuning the listed parameters for each individual two-qubit rotation. Note that for the really small gates (4 ns and smaller), this protocol starts to break down, thus we do not see a similar pattern in parameters choices determined by the numerical exploration.

\vspace{0.5cm}

\begin{table}
\begin{ruledtabular}
	\begin{tabular}{cc|cccc|cccc}
		\multicolumn{10}{c}{\textbf{$i$SWAP waveforms}}\\
		\thiccboi{2pt}\\[-1.8ex]
		\multicolumn{10}{c}{Cosine profile---Eq.~\ref{eq:envelopes}a}\\
		\colrule
		& & \multicolumn{4}{c|}{Non positive definite ($\alpha=2$)} & \multicolumn{4}{c}{Positive definite ($\alpha = 1$)} \\
		\colrule
		Fraction & $t_g$ (ns) & $A$ (MHz) & \multicolumn{2}{c}{$f$ (MHz)} & Error $(10^{-5})$ & $A$ (MHz) & \multicolumn{2}{c}{$f$ (MHz)} & Error $(10^{-5})$\\
		\colrule
		$1$ & 36 & 13.91 & \multicolumn{2}{c}{-515.31} & 0.51 & 13.89 & \multicolumn{2}{c}{-377.61} & 5.93 \\
		$3/4$ & 27 & 13.91 & \multicolumn{2}{c}{-524.11} & 0.23 & 13.86 & \multicolumn{2}{c}{-374.14} & 0.44 \\
		$1/2$ & 18 & 13.87 & \multicolumn{2}{c}{-529.38} & 0.20 & 13.85 & \multicolumn{2}{c}{-410.17} & 14.67 \\
		$1/4$ & 9 & 13.84 & \multicolumn{2}{c}{-528.47} & 1.10 & 13.74 & \multicolumn{2}{c}{-528.06} & 17.98 \\
		$1/5$ & 6 & 13.77 & \multicolumn{2}{c}{-642.43} & 8.99 & 13.66 & \multicolumn{2}{c}{-644.62} & 16.67 \\
		$1/8$ & 4.5 & 13.04 & \multicolumn{2}{c}{-571.80} & 20.46 & 13.38 & \multicolumn{2}{c}{-548.56} & 8.46 \\
		$1/12$ & 3 & 10.25 & \multicolumn{2}{c}{-784.42} & 97.44 & 40.00 & \multicolumn{2}{c}{-183.81} & 113.59 \\
		\thiccboi{2pt}\\[-1.8ex]
		\multicolumn{10}{c}{Tangent profile---Eq.~\ref{eq:envelopes}b}\\
		\colrule
		Fraction & $t_g$ (ns) & $A$ (MHz) & $f$ (MHz) & $\gamma$ & Error $(10^{-5})$ & $A$ (MHz) & $f$ (MHz) & $\gamma$ & Error $(10^{-5})$\\
		\colrule
		$1$ & 36 & 8.82 & -523.28 & 9.37 & 0.36 & 8.83 & -369.09 & 9.35 & 0.49 \\
		$3/4$ & 27 & 10.17 & -525.38 & 6.32 & 0.45 & 8.06 & -384.34 & 13.99 & 4.67 \\
		$1/2$ & 18 & 10.59 & -525.40 & 5.74 & 0.56 & 8.16 & -363.04 & 12.67 & 1.63 \\
		$1/4$ & 9 & 9.94 & -591.50 & 6.40 & 1.74 & 10.23 & -458.40 & 6.11 & 7.09 \\
		$1/5$ & 6 & 14.92 & -576.66 & 3.62 & 0.58 & 14.85 & -565.25 & 3.63 & 7.58 \\
		$1/8$ & 4.5 & 9.58 & -593.94 & 9.06 & 0.51 & 9.50 & -603.31 & 7.79 & 3.87 \\
		$1/12$ & 3 & 6.94 & -684.12 & 15.00 & 4.60 & 40.00 & -167.78 & 6.87 & 1.11 \\
	\end{tabular}
\end{ruledtabular}
\caption{\label{tab:iSWAP-waveforms}
Parameters for $i\text{SWAP}$ waveform results in Fig.~\ref{fig:edg_pulses} and error rates in Fig.~\ref{fig:edg_results}.}
\end{table}


\begin{table}
	\begin{ruledtabular}
		\begin{tabular}{cc|cccc|cccc}
			\multicolumn{10}{c}{\textbf{$XX$ waveforms}}\\
			\thiccboi{2pt}\\[-1.8ex]
			\multicolumn{10}{c}{Cosine profile---Eq.~\ref{eq:envelopes}a}\\
			\colrule
			& & \multicolumn{4}{c|}{Non positive definite ($\alpha=2$)} & \multicolumn{4}{c}{Positive 	definite ($\alpha = 1$)} \\
			\colrule
			Fraction & $t_g$ (ns) & $A$ (MHz) & \multicolumn{2}{c}{$f$ (MHz)} & Error $(10^{-5})$ & 	$A$ (MHz) & \multicolumn{2}{c}{$f$ (MHz)} & Error $(10^{-5})$\\
			\colrule
			$1$ & 24 & 10.42 & \multicolumn{2}{c}{-854.11} & 7.11 & 10.42 & \multicolumn{2}{c}{-698.	30} & 20.21 \\
			$3/4$ & 18 & 10.42 & \multicolumn{2}{c}{-861.22} & 4.42 & 10.41 & \multicolumn{2}{c}{-717.	74} & 16.50 \\
			$1/2$ & 12 & 10.41 & \multicolumn{2}{c}{-850.74} & 2.88 & 10.39 & \multicolumn{2}{c}{-768.	57} & 11.88 \\
			$1/4$ & 6 & 10.32 & \multicolumn{2}{c}{-955.75} & 11.16 & 10.31 & \multicolumn{2}{c}{-961.	83} & 14.13 \\
			$1/5$ & 4 & 8.33 & \multicolumn{2}{c}{0.00} & 172.07 & 8.30 & \multicolumn{2}{c}{-923.54} 	& 171.03 \\
			$1/8$ & 3 & 6.27 & \multicolumn{2}{c}{0.00} & 191.99 & 40.00 & \multicolumn{2}{c}{-171.60} 	& 133.28 \\
			$1/12$ & 2 & 5.00 & \multicolumn{2}{c}{-9.06} & 123.77 & 40.00 & \multicolumn{2}{c}{-251.	10} & 83.48 \\
			\thiccboi{2pt}\\[-1.8ex]
			\multicolumn{10}{c}{Tangent profile---Eq.~\ref{eq:envelopes}b}\\
			\colrule
			Fraction & $t_g$ (ns) & $A$ (MHz) & $f$ (MHz) & $\gamma$ & Error $(10^{-5})$ & $A$ (MHz) & 	$f$ (MHz) & $\gamma$ & Error $(10^{-5})$\\
			\colrule
			$1$ & 24 & 7.66 & -851.68 & 6.25 & 5.20 & 8.28 & -715.89 & 5.38 & 20.31 \\
			$3/4$ & 18 & 10.08 & -870.65 & 4.11 & 4.03 & 8.12 & -752.41 & 5.56 & 16.74 \\
			$1/2$ & 12 & 6.62 & -898.74 & 9.29 & 4.33 & 6.56 & -801.57 & 9.52 & 9.59 \\
			$1/4$ & 6 & 12.52 & -901.87 & 3.34 & 1.92 & 12.98 & -909.24 & 3.24 & 5.21 \\
			$1/5$ & 4 & 6.34 & -892.32 & 13.41 & 2.38 & 6.24 & -903.65 & 13.65 & 2.29 \\
			$1/8$ & 3 & 5.18 & 0.00 & 15.00 & 66.43 & 5.18 & 0.00 & 15.00 & 66.43 \\
			$1/12$ & 2 & 5.00 & -20.70 & 15.00 & 93.95 & 18.28 & -250.01 & 8.09 & 79.21 \\
		\end{tabular}
	\end{ruledtabular}
	\caption{\label{tab:xx-waveforms}
	Parameters for $XX$ waveform results in Fig.~\ref{fig:edg_pulses} and error rates in Fig.~\ref{fig:edg_results}.}
\end{table}

\vspace{0.5cm}

We note that it would be convenient for a realization of this technique to use the same set of parameters for all partial rotations of a gate by tuning more parameters. For example, using device parameters for the current experimental realization---qubit nonlinearities $-0.16346\times2\pi$ GHz and $-0.254655 \times 2\pi$ GHz---and ignoring qubit energies in the rotating frame, if we modify Eq.~\ref{eq:couplingStrength} to $\Omega(t) = \Omega_0(t)\left[c + \alpha\sin(2\pi f)\right]$ such that $c$ and $\alpha$ can be arbitrarily tuned, using Eq.~\ref{eq:envelopes}a we can numerically find $\{ A, f, \alpha, c \} = \{ 82.645\ \text{MHz}, 1.868\ \text{GHz}, 1.333, 0.336 \}$ to get the gate errors $ \{1.21, 0.67, 0.30, 0.49, 0.29, 7.43, 102.99\}\times 10^{-6} $, shown in the figure below.

\begin{figure}\label{fig:single_set}
	\includegraphics[width = 0.95\textwidth]{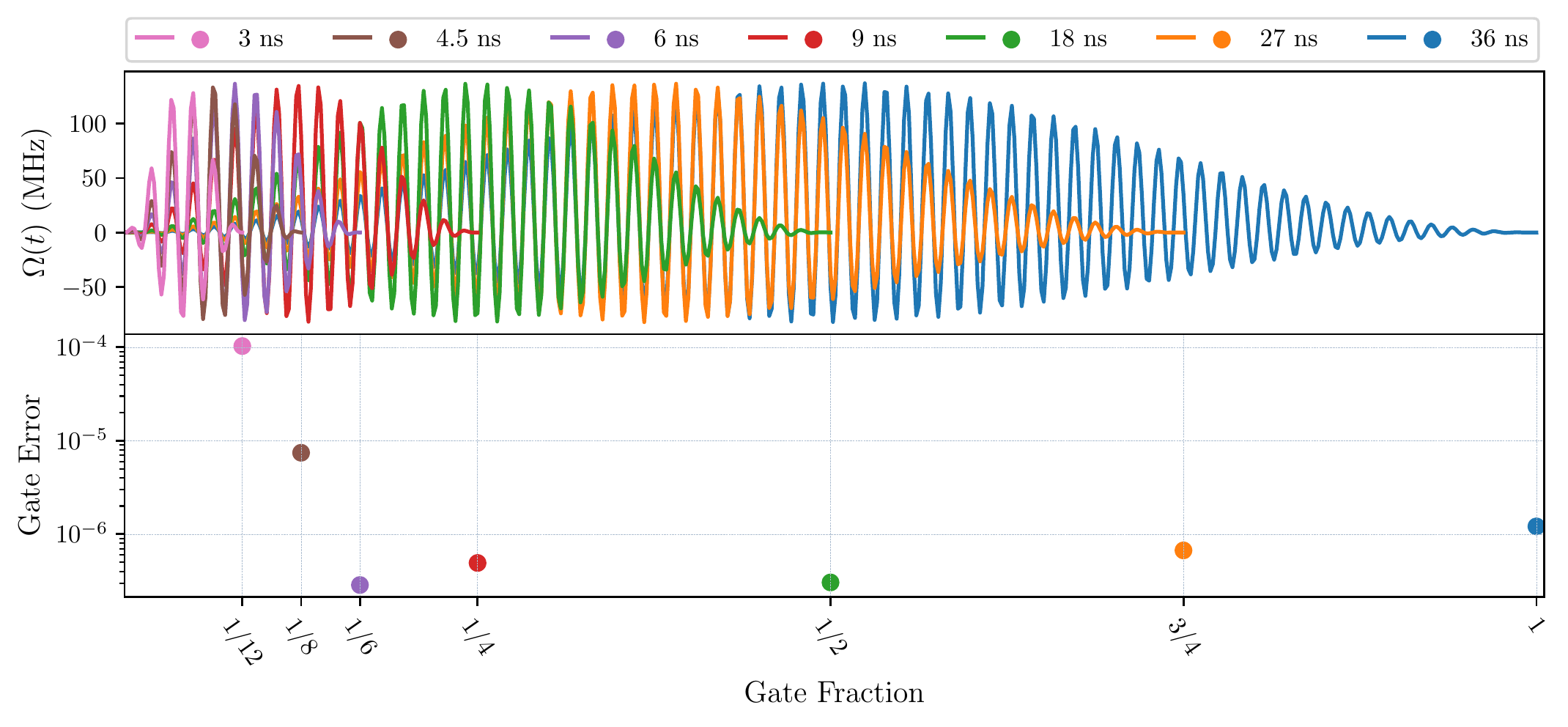}
	\caption{Fractional gates using the same set of parameters. $\{ A, f, \alpha, c \} = \{ 82.645\ \text{MHz}, 1.868\ \text{GHz}, 1.333, 0.336 \}$ (top) to get the gate errors $ \{1.21, 0.67, 0.30, 0.49, 0.29, 7.43, 102.99\}\times 10^{-6} $(bottom).}
\end{figure}

\begin{table}
		\begin{tabular}{c|c|c|c}
			Number of qubits & Stock & Continuous & Error-divisible \\
			\thiccboi{2pt}
			4 & 0.507243 & 0.575992 & 0.817031 \\
			 6 &  0.509382 & 0.585032 & 0.802391 \\
			 8 &  0.512273 & 0.59149 & 0.770956 \\
			 10 &  0.514339 & 0.59568 & 0.757204 \\
			 12 &  0.515807 & 0.598545 & 0.754664 \\
			 \thiccboi{2pt}
		\end{tabular}
	\caption{\label{tab:vqe-results}
	Data fir results plotted in Fig.~\ref{fig:vqe_results}, comparing the percent of the ground state energy that each type of gate set is able to produce.}
\end{table}

%% file: appendix/circuits.tex
\section{Reference implementations}\label{app:circuits}

We use the following notation:

\begin{center}
	\includegraphics{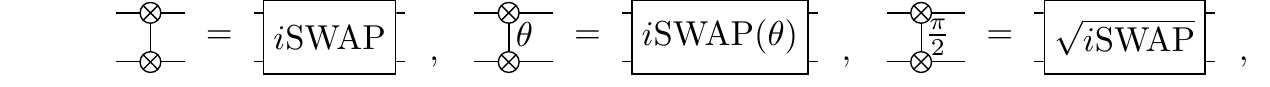}
\end{center}
where $i\text{SWAP}(\theta) = \exp\left[i\theta/2\left(\sigma_1^+\sigma_2^- + \sigma_1^-\sigma_2^+\right)\right]$, with a full $i\text{SWAP} = i\text{SWAP}(\pi)$, and $\sqrt{i\text{SWAP}} = i\text{SWAP}(\pi/2)$. Using a stock native gate set---with CNOT as the base two-qubit gate, we can implement the two-qubit SWAP and $i$SWAP gates.

\begin{center}
	\includegraphics{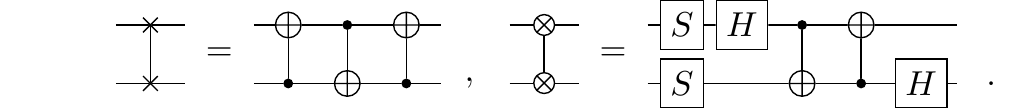}
\end{center}
With access to a broader set of native two-qubit gates, arbitrary two-qubit gates can be implemented more efficiently. For example, with access to a native $\sqrt{i\text{SWAP}}$, we can achieve the lowest-error implementation of SWAP that the authors are aware of from \cite{lebedev_extended_2018},

\begin{center}
	\includegraphics{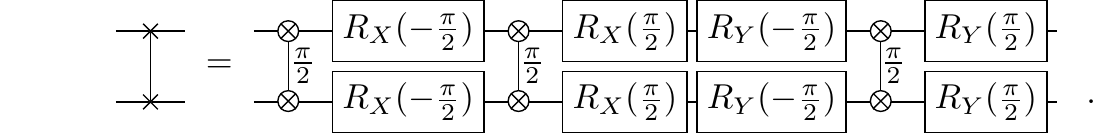}
\end{center}
This explicitly demonstrates the benefit of having access to a native partial gate like $\sqrt{i\text{SWAP}}$ for implementation of other two-qubit gates. More broadly, any arbitrary $i\text{SWAP}(\theta)$ rotation can be implemented using native $\sqrt{i\text{SWAP}}$ gates,

\begin{center}
	\includegraphics{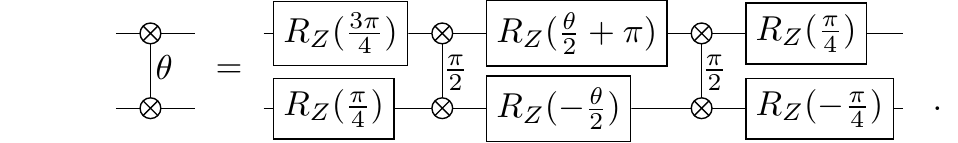}
\end{center}
This decomposition highlights the benefit of having access to a native set of arbitrary $\theta$ rotations that would allow the reduction of two-qubit gate error proliferation, as well as energy loss by virtue of spending less time on a single gate.